\documentclass[conference]{IEEEtran}

\usepackage[utf8]{inputenc}
\usepackage{cite}

\usepackage{amsmath}
\makeatletter
\@ifundefined{Bbbk}{}{}
\makeatother
\usepackage{amssymb}
\usepackage{array}
\usepackage{multirow}
\usepackage{graphicx}
\usepackage{balance}
\usepackage{booktabs}
\usepackage{tabularx}
\usepackage{makecell}
\usepackage[caption=false,font=footnotesize]{subfig}
\usepackage{wrapfig}
\usepackage{cuted}
\usepackage{doi}
\usepackage{orcidlink}

\usepackage{nccmath}
\usepackage[most]{tcolorbox}
\usepackage{anyfontsize}
\usepackage[all]{nowidow}
\usepackage[normalem]{ulem}
\usepackage{xcolor}
\usepackage{colortbl}
\usepackage{enumitem}

\usepackage{algorithm}
\usepackage{algpseudocode}
\usepackage{listings}

\usepackage{url}

\usepackage{tikz}
\usepackage[framemethod=TikZ]{mdframed}
\definecolor{light-gray}{gray}{0.9}
\mdfsetup{skipabove=5pt,skipbelow=3pt}
\mdfdefinestyle{MyFrame}{linecolor=black,
  outerlinewidth=0.15pt,
  roundcorner=3pt,
  innertopmargin=2pt,
  innerbottommargin=2pt,
  innerrightmargin=4pt,
  innerleftmargin=4pt,
  backgroundcolor=light-gray}
\mdfdefinestyle{RQFrame}{linecolor=black,
  outerlinewidth=0.15pt,
  roundcorner=3pt,
  innertopmargin=2pt,
  innerbottommargin=2pt,
  innerrightmargin=4pt,
  innerleftmargin=4pt,
  backgroundcolor=light-gray}

\usepackage{pifont}
\newcommand{\cmark}{\textcolor{green!60!black}{\checkmark}}
\newcommand{\xmark}{\textcolor{red!70!black}{\(\times\)}}
\newcommand{\pmark}{\textcolor{orange!80!black}{\(\triangleright\)}}

\newcolumntype{Y}{>{\centering\arraybackslash}X}

\definecolor{commentGray}{RGB}{120,120,120}
\renewcommand{\algorithmiccomment}[1]{\bgroup\color{commentGray}{//#1}\egroup}

\newcommand{\name}{LMVQA}

\newcommand\resq[1]{\noindent
\fcolorbox{green!40!black}{green!5}{\parbox{0.95\columnwidth}{\noindent #1}}\\
}

\definecolor{javagreen}{rgb}{0.25,0.5,0.35}
\lstdefinestyle{Alg}{
  basicstyle=\ttfamily\footnotesize,
  breaklines=true,
  tabsize=2,
  mathescape,
  numbers=left,
  xleftmargin=2.5em,
  xrightmargin=0.5em,
  frame=tb,
  framexleftmargin=2em,
  emph={Algorithm,Input,Output,for,each,do,if,else,Function,while,let,be,repeat,until,return,times,and,or,break,in,then,},
  emphstyle={\textbf},
  escapechar=?,
  morecomment=[l][\color{javagreen}]{//},
  columns=flexible,
}


\begin{document}

\title{Question Answering for Diagram-Rich \\ Technical Meeting Videos}

\author{ \IEEEauthorblockN{ Zhuoran Xu\IEEEauthorrefmark{1}\IEEEauthorrefmark{2}, Jia Li\IEEEauthorrefmark{1}\orcidlink{0009-0000-8859-4541}\IEEEauthorrefmark{2}, Dayuan Tan\IEEEauthorrefmark{2}, Mark Cole\IEEEauthorrefmark{2}, Ish Ashraf\IEEEauthorrefmark{2}, Sandeep Puri\IEEEauthorrefmark{2}, \\Mehrdad Sabetzadeh\IEEEauthorrefmark{1}\orcidlink{0000-0002-4711-8319}, Shiva Nejati\IEEEauthorrefmark{1}\orcidlink{0000-0002-0281-8231}}
\IEEEauthorblockA{\IEEEauthorrefmark{1}University of Ottawa, Ottawa, ON, Canada} \IEEEauthorblockA{\IEEEauthorrefmark{2}Ciena Corp, Ottawa, ON, Canada} \IEEEauthorblockA{ \{zxu045, jli714, m.sabetzadeh, snejati\}@uottawa.ca \\ \{datan, mcole, iashraf, spuri\}@ciena.com} }

\maketitle

\begin{abstract}
Software engineering increasingly relies on asynchronous communication artifacts, including recorded meetings where stakeholders discuss concerns, rationale, and decisions. These meetings often include diagram-based representations of requirements, system behavior, component interactions, and trace dependencies. Accessing knowledge from these meetings is challenging because recordings are long and relevant evidence is distributed across speech, slides, and technical diagrams. This paper reports our industrial experience developing and evaluating \name, an LLM-based multimodal question-answering system for technical meeting videos. Developed in collaboration with engineers at Ciena, \name\ supports the understanding of requirements and design intent by grounding answers in audio and visual evidence, with explicit handling of diagram-rich content such as requirements and UML diagrams. It processes each video once to build a reusable time-stamped evidence corpus for grounded question answering. Across a Ciena dataset and a public dataset, we show that \name\ significantly improves answer accuracy compared to a state-of-the-art baseline, from 31\% to 94\% on the Ciena dataset and from 21\% to 88\% on the public dataset, with larger gains on diagram-rich videos. We further show that, after one-time indexing, \name\ reduces average response time from 81.3s to 3.3s on Ciena and from 98.4s to 9.2s on the public dataset, while lowering average token-based LLM API cost by about 75\%.  Finally, our interviews with three domain experts show that engineers particularly value \name\ for locating software-engineering-relevant information, revisiting rationale, and tracing answers to specific video segments.
\end{abstract}


\section{Introduction}
\label{sec:intro}
A shared understanding of system goals, assumptions, constraints, design decisions, and rationale is essential to successful software projects~\cite{Lamsweerde:00}. Misalignment among stakeholders can create conflicting interpretations of requirements and design decisions, reducing efficiency and increasing project risk~\cite{Lloyd:02}. This challenge is amplified in industrial settings, where distributed stakeholders cannot always participate in synchronous discussions. As a result, asynchronous communication has become increasingly important in software engineering~\cite{Girgensohn:15}.

Recorded technical meetings are one such medium~\cite{Nagel:RE24VisionVideos}. They often capture requirements- and design-relevant information, including stakeholder concerns, rationale, decisions, dependencies, and visual explanations of system behavior, architecture, workflows, and design alternatives. Engineers revisit these videos to clarify requirements, recover decision rationale, and understand component dependencies. These activities are central to software maintenance and evolution, where engineers must frequently retrieve historical context, rationale, and architectural knowledge from prior technical discussions. Prior work shows that recorded video supports knowledge transfer when synchronous participation is not possible~\cite{Skylar:09,Palsole:08}. However, these recordings are difficult to use efficiently because they are long, users often have specific questions, and relevant evidence is distributed across speech, slides, and technical diagrams.

This challenge is especially important for software-intensive systems because technical meeting videos are highly multimodal. They combine spoken explanations with slides, source code, and structured diagrams such as domain, goal, UML, data-flow, and process models~\cite{Ponzanelli:ICSE16CodeTube}. These diagrams capture entities, relationships, hierarchies, interactions, and dependencies central to requirements understanding and design intent. Although prior work supports video search through text-based tagging~\cite{EscobarAvila:ICSEC17VideoTagging,Parra:2018AutoTagVideoTutorials} and code extraction from screencasts~\cite{Bao:TOSEM20psc2code,Bao:TSE19VTRevolution,Malkadi:ASE23CodeExtraction,Yang:22}, precise question answering over long technical meeting videos remains difficult when evidence is diagrammatic and dispersed across modalities~\cite{Yan:ICSE24VideoBugDuplicate}.

Large Language Models (LLMs) are increasingly used to build question-answering systems for software engineering tasks~\cite{khamsepour:26,Chaudhary:24,Abedu:26,Yang:25}. Multimodal LLMs further make it possible to process image and audio content in addition to text. However, existing video question-answering approaches are not well suited to software engineering questions over long technical recordings. Many are designed for short videos and do not scale well to hours-long content~\cite{flamingo,video-llama,video-llava,video-chatgpt}. Some process visual information only coarsely~\cite{Ma:25,zhang:24}, making them less effective for diagram-intensive material, while others do not fully integrate the audio and visual evidence available in videos~\cite{flamingo,video-chatgpt,video-llava,longvlm,Ma:25}. These limitations matter in fields like software engineering, where users often need grounded answers that connect spoken discussion with diagrammatic evidence. To the best of our knowledge, there is no prior work on handling open-ended question answering over long technical meeting videos where answers require combining spoken discussions with software diagrams.

To address this challenge, we propose \textit{\name}, an \textbf{L}LM-based \textbf{M}ultimodal system for technical \textbf{V}ideos \textbf{Q}uestion-\textbf{A}nswering. \name\ processes each video once to build a reusable, time-stamped evidence corpus from both visual and audio streams. On the visual side, it samples representative frames, detects diagram-containing frames, and applies diagram-type-aware extraction for common software engineering diagrams, including domain models  and UML diagrams. On the audio side, it transcribes speech into time-stamped segments. These multimodal evidence units are then embedded and stored in a vector database for retrieval. Provided with a question, \name\ retrieves the most relevant evidence units and uses them as the context for answer generation. This design
supports long videos by processing them into reusable indexed chunks and explicitly targets the diagram-centric nature of technical meeting content.

We evaluate \name\ on two technical meeting video question-answering datasets: an industrial dataset from Ciena and a public dataset derived from undergraduate software engineering lectures~\cite{loviqa-prompts}. Compared with a state-of-the-art baseline,  DrVideo~\cite{Ma:25}, \name\ improves average accuracy from 31\% to 94\% on the Ciena dataset and from 21\% to 88\% on the public dataset, with especially strong gains on diagram-based questions. \name\ also improves efficiency: after one-time video indexing, it reduces average per-question latency from 81.3s to 3.3s on Ciena and from 98.4s to 9.2s on the public dataset, while lowering average LLM API cost by approximately 75\%. Interviews with three Ciena engineers further indicate that \name\ helps users identify  relevant video segments, quickly find specific information without rewatching the entire recording, verify answers using time-stamped evidence, and interpret diagram-related content in technical videos.

\noindent\textbf{Contributions.} This paper makes the following contributions:

\textbf{-} We present a software-engineering-aware multimodal question-answering approach for long technical meeting videos. Our approach combines time-stamped evidence construction with diagram-type-aware extraction, enabling grounded answers to questions about diagram-rich technical discussions.

\textbf{-} We implement our approach in \name, a two-stage multimodal system that constructs a reusable evidence corpus and answers queries via retrieval, with explicit support for diagram detection and diagram-type-aware extraction.

\textbf{-} We evaluate \name\ on an industrial dataset from Ciena and a public course dataset related to software  engineering, showing substantial accuracy improvements over a state-of-the-art baseline, DrVideo~\cite{Ma:25}, while lowering overall cost through corpus reuse. We further conduct qualitative interviews with three Ciena engineers, who reported that \name\ helps users efficiently locate, verify, and interpret information in long technical meeting videos.

 \section{Industry Context}
Large technology companies such as Ciena are exploring AI solutions, including chatbots, to automate routine work and improve information retrieval across their complex technical content portfolios~\cite{Abedu:26,Chaudhary:24,khamsepour:26}. One particularly challenging content type is archived technical meeting videos, which are recorded for software engineering purposes. For example, these recordings may include stakeholder interview and design meetings for requirements elicitation and design exploration, capturing user needs, design alternatives, priorities, constraints, and rationale.
They also help preserve knowledge when experienced engineers leave or transition roles, allowing recorded explanations to serve as institutional memory.
In practice, these recordings often become part of the long-term institutional memory needed to support onboarding, system evolution, and maintenance activities across teams. In addition, they enable cross-team communication by helping distributed teams stay aligned with evolving requirements, designs, constraints, and technical decisions.

However, retrieving knowledge from these video archives is challenging. Ciena’s technical meetings typically span 45–90 minutes and combine spoken discussion with slides, demos, and technical diagrams (e.g., UML, domain models, and data flows). Engineers often need answers to very specific questions, such as ``What decision was made regarding the user authentication requirement discussed in last week’s meeting?'' or ``What design constraints were identified for the XYZ component during our design brainstorming meeting?'' Finding such specific information can be slow and frustrating: it typically requires watching large portions of a long recording or relying on coarse, timestamp-based navigation that is often too imprecise to pinpoint the exact moment where the relevant detail was discussed or shown. Another key challenge is that much of the critical information is visual. In our analysis, 24\% of the video frames from Ciena's technical meetings include diagrams whose semantics cannot be recovered from speech-to-text alone. Existing video search tools rely largely on metadata and transcripts, so they fail when relevant details appear in slides or diagrams. As Ciena produces hundreds of hours of recordings each year, scalable retrieval requires automated methods that jointly index and query both audio and visual content~\cite{Yan:ICSE24VideoBugDuplicate,Yang:22}.

Our goal in this paper is to develop an LLM-based question-answering system for technical meeting videos containing rich diagrammatic content. In particular, our question-answering system aims to address the following needs, identified through our discussions with Ciena's engineering teams:  \textbf{(1)~Diagram understanding:} The question-answering system must accurately interpret and answer questions about technical diagrams, including their entities, relationships, and hierarchical structures. \textbf{(2)~Multimodal fusion:} Answers should integrate information from both visual content (slides, diagrams, demonstrations) and audio content (spoken explanations, discussions). \textbf{(3)~Factual grounding:} Generated answers must be grounded in the video content, with the ability to trace answers back to specific segments when needed. This helps reduce hallucination risks and ensures that stakeholders can verify the information. \textbf{(4)~Cost efficiency:} Given the volume of videos to be processed, the system must be cost-effective for enterprise deployment. Commercial LLMs with token-based API pricing can become expensive when processing long-form content. \textbf{(5)~Reasonable response time:} Users expect interactive response times, particularly for frequently queried videos. A system that requires hours to process a video before it can be queried would not meet  usability expectations.

 \section{Our Approach (\name)}
\label{section:entire-chatbot}
\begin{figure}
    \centering
\includegraphics[width=0.95\linewidth]{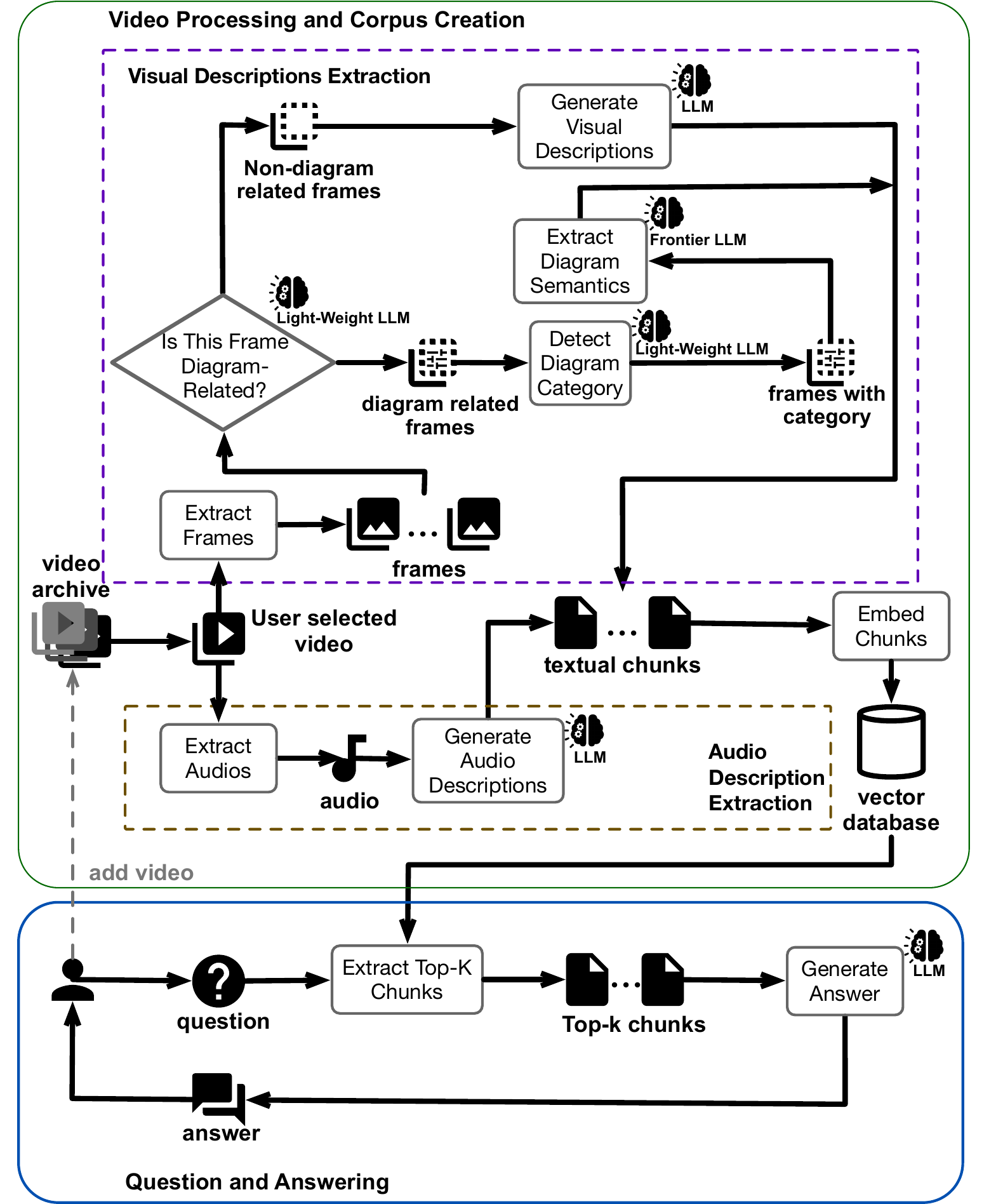}
    \caption{Overview of our LLM-based multimodal approach for technical video question answering (\name)}
    \label{fig:overview}
    \vspace*{-.2cm}
\end{figure}

Figure~\ref{fig:overview} shows an overview of \name\ which consists of two stages: (1)~video processing and corpus creation, and (2)~question answering. In the first stage, \name\ converts each input video into a collection of time-stamped textual chunks derived from both the visual and audio streams, and stores their embeddings in a vector database for retrieval. In the second stage, given a user query, \name\ retrieves the most relevant chunks and uses them as evidence to generate an answer. Section~\ref{sec:video-preprocessing} explains how \name\ constructs a time-stamped chunk corpus from the visual and audio streams and stores it in a vector database. Section~\ref{sec:qa} then describes retrieval and answer generation.

\subsection{Video Processing and Corpus Creation}
\label{sec:video-preprocessing}

For each input video, \name\ processes the visual and audio streams separately and converts them into time-stamped chunks through the following two steps: 

\noindent\textbf{(1)~Visual description extraction.} Technical meeting videos often contain many identical or near-identical consecutive frames. For instance,  presentation slides can remain unchanged for several minutes and meeting discussions frequently show fixed participant windows with minimal variation. As a result, a high proportion of frames of technical meeting videos are duplicate. For example, an online video recorded at the standard frame rate of $30$ FPS~\cite{framerates:25} generates approximately $162{,}000$ frames in a 90-minute session, most of which contain little additional semantic information. 

To reduce redundancy, \name\ samples technical meeting videos at a low rate (e.g., $1$ FPS) and selects a subset of the sampled frames using the Structural Similarity Index (SSIM)~\cite{wang:04}. Specifically, a sampled frame is retained as a \emph{keyframe} if its SSIM score with respect to the most recently retained keyframe falls below a predefined threshold, indicating a meaningful visual change. Each keyframe is then assigned a time interval corresponding to the segment of the video for which it is visually representative.

\begin{figure}[t]
    \centering
    \includegraphics[width=0.95\linewidth]{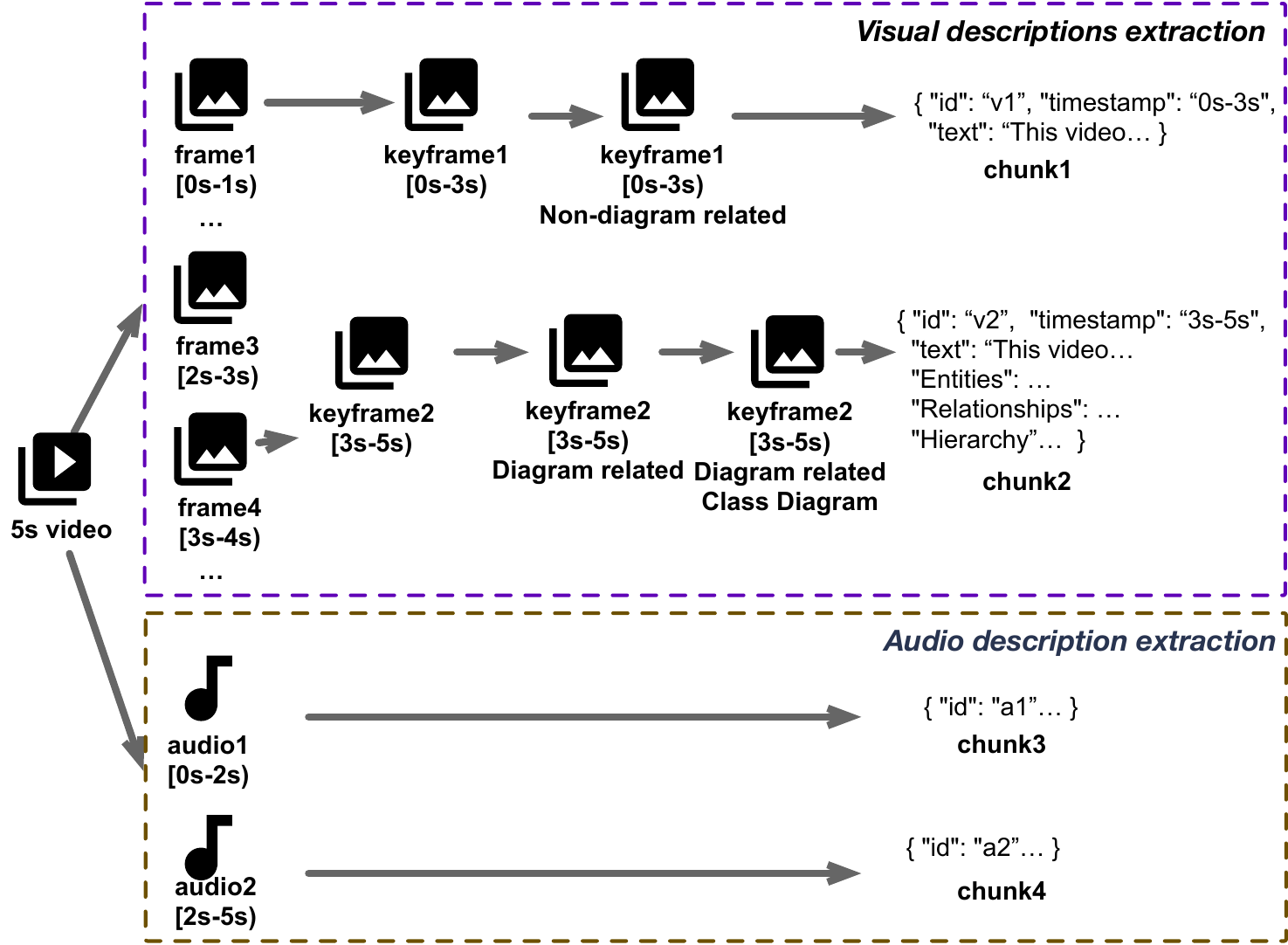}
    \caption{Excerpt of an example showing how visual and audio content from a 5-second video is segmented into chunks.}
    \label{fig:chunkexamples}
     \vspace*{-.45cm}
\end{figure}

Figure~\ref{fig:chunkexamples} illustrates the keyframe extraction process. From a 5-second video, we sample one frame per second, yielding five frames. Using SSIM, each new frame is compared with the last retained keyframe and kept only if it exhibits a significant visual difference. In this example, two keyframes are retained, denoted as keyframe1 and keyframe2 in the figure. The interval assigned to keyframe1 is $[0\mathrm{s},3\mathrm{s})$, while the interval assigned to keyframe2 is $[3\mathrm{s},5\mathrm{s})$. Together, these two frames serve as the visual representation of the 5-second video.

Once keyframes are selected, \name\ identifies whether each frame contains a diagram. Since diagram-related frames encode richer semantics than non-diagram frames, \name\ first uses a lightweight LLM to classify each keyframe as diagram-related or non-diagram, where diagram-related means the frame contains at least one diagram. This classification follows the diagram-detection prompt in Appendix~\ref{appendix:prompt-outlines}, Listing a. \name\ then uses a lightweight LLM to generate concise visual descriptions for non-diagram frames, following the non-diagram captioning prompt in Listing b.

To handle diagrams with different syntax and semantics, \name\ uses the same lightweight LLM to classify each diagram-related keyframe into one of the following categories: four UML types (class, sequence, state/activity, and use-case), network/topology diagrams, architecture/workflow diagrams, and an ``other'' category. This classification follows the diagram classification prompt in Appendix~\ref{appendix:prompt-outlines}, Listing c. The diagram categories and their category-specific extraction fields are summarized in Listing d. \name\ then processes each diagram-related frame with a frontier LLM using the diagram frame extraction prompt in Listing e, together with the corresponding category-specific extraction fields. The prompts in Listings a to d in the appendix use Chain-of-Thought prompting~\cite{wei:22} and few-shot In-Context Learning~\cite{Brown:20} to better capture relationships among diagram elements through example diagram frames paired with detailed semantic descriptions. The actual prompts are available online~\cite{loviqa-prompts}.

The descriptions of the keyframes, whether or not they contain diagrams, obtained by LLMs form a set of chunks. Each chunk corresponding to a keyframe is encoded in JSON format and includes a unique chunk ID, a textual description of the keyframe, the keyframe’s time interval, and the term ``visual'', indicating that the chunk captures the content of a visual keyframe. For example, as illustrated in Figure~\ref{fig:chunkexamples}, two textual chunks, denoted as chunk1 and chunk2, are generated for the two keyframes, keyframe1 and keyframe2.

\noindent\textbf{(2)~Audio description extraction.} In parallel with visual processing, \name\ transcribes the input video’s audio track using an automatic speech recognition (ASR) system with voice activity detection enabled (e.g., Whisper~\cite{radford2023whisper}). The resulting transcript is segmented into sentence-level units. To reduce noise while preserving meaning, we restore punctuation and remove common filler words, such as ``um'', ``uh'' and ``you know''. \name\ then converts the audio track into chunks, with each chunk corresponding to one sentence. Each audio chunk is assigned the time interval of its corresponding sentence and is encoded in JSON format with a unique chunk ID, the sentence text, its time interval, and the label ``audio'' to indicate that it originates from the audio track. The audio track in Figure~\ref{fig:chunkexamples} contains two sentences, and two chunks, chunk3 and chunk4, are derived from it.

All chunks generated from a single video, whether derived from visual keyframes or audio sentences, are aggregated and serialized into a corpus. The corpus is then embedded in a high-dimensional vector space using a pretrained embedding model, and the resulting embeddings are stored in a vector database for efficient retrieval.

\subsection{Question Answering}
\label{sec:qa}

Given a user query, \name\ first encodes it into the same vector space as the corpus. It then retrieves the top-$k$ most relevant chunks from the vector database using cosine similarity~\cite{salton:75}. The choice of $k$ involves a trade-off: a large $k$ increases computational cost, while a small $k$ may exclude relevant information. In our setting, retrieving the top-$k$ chunks provides a practical balance between answer quality and efficiency by supplying the LLM with sufficient evidence while keeping the prompt size manageable.

\name\ converts the encoded query and retrieved chunks into a structured prompt for answer generation, following the RAG-based video question-answering prompt in Appendix~\ref{appendix:prompt-outlines}, Listing f. This prompt enforces two key instructions: (i)~the LLM must rely exclusively on the retrieved chunks, and (ii) the LLM must explicitly abstain from speculation when sufficient evidence is absent. To construct the prompt, we place the user query first, append the retrieved top-ranked chunks as supporting evidence, and then add explicit instructions that reduce hallucinations by requiring the LLM to avoid answers unsupported by the retrieved evidence. When the retrieved evidence includes diagram-related frames, the prompt further incorporates chain-of-thought reasoning and few-shot examples to enhance the completeness and accuracy of diagram-based answers. In addition to generating an answer, \name\ returns the top-$k$ most relevant chunks together with their corresponding time intervals and content summaries, enabling users to trace the answer back to its original locations in the video for traceability and verification.

 \section{Empirical Evaluation}
\label{sec:eval}

In this section, we evaluate  \name\ based on the following research questions (RQs):

\textbf{RQ1 (Accuracy).} \textit{Can \name\ accurately answer questions based on technical meeting videos?} To address RQ1, we compare the accuracy of \name\ with that of a state-of-the-art video processing method,  DrVideo~\cite{Ma:25},  across two video datasets: One based on technical meeting recordings at  Ciena, and the other  derived from lectures of two undergraduate software engineering courses at the University of Ottawa.

\textbf{RQ2 (Efficiency).} \textit{How efficiently can \name\ answer questions  based on technical meeting videos?} We evaluate \name's response time and cost, and compare its performance with the baseline (DrVideo) on both datasets.

\textbf{RQ3 (Practitioner Feedback). }\textit{How do engineers perceive \name?} To answer RQ3, we invite three practicing engineers from Ciena, who are not co-authors of this paper,  to use \name, and collect their feedback on the perceived usefulness, accuracy, and evidence-tracing support of \name.

\subsection{Dataset}
We evaluate \name\ on two datasets. The first, provided by Ciena, consists of meeting recordings and project introductions (hereafter referred to as the ``Ciena" dataset). The second, which we constructed, comprises lecture recordings from two undergraduate software engineering courses (hereafter referred to as the ``Course'' dataset). The Course dataset is publicly available online~\cite{loviqa-prompts}. Table~\ref{tab:datasets} summarizes the main characteristics of the two datasets, including the number of videos, their average durations and sizes, the average number of chunks per video, and the total number of questions. The following paragraphs describe the datasets in more detail.

\noindent\textbf{Ciena dataset.} This dataset contains five videos contributed by four different internal groups at Ciena. The videos include work-progress presentations that summarize project context, stakeholder requirements, design choices, and operational workflows. 
In total, the dataset includes $236$ questions ($47.2$ questions per video). The questions were provided by Ciena and were derived from two sources: (a) audience questions asked during the presentations (capturing clarifications and ``why/how'' reasoning), and (b) \emph{knowledge-check} questions about the key concepts employees were expected to learn from the recordings. Ciena engineers developed the ground-truth answers to the questions.

\noindent\textbf{Course dataset.} This dataset is made up of five lecture recordings from two undergraduate software engineering courses, totaling $488.6$~MB and $363.02$ minutes, with an average duration of $72.6$ minutes per video. One course focuses on general software engineering concepts (e.g., requirements and domain analysis, architectural and design modeling), and the other focuses on Java programming (e.g., object-oriented design and implementation). Both courses make extensive use of UML. The questions and ground-truth answers were created by two university students, neither of whom is a co-author of this paper. The questions were formulated based on the recordings to ensure coverage of both verbal explanations and visually grounded information. The questions span (a)~conceptual understanding of software  engineering topics introduced by the lecturer, (b)~code- and API-level reasoning in Java (e.g., inheritance, class responsibilities, and method behavior), and (c)~diagram-centric interpretation of UML elements (e.g., identifying diagram components, relationships, and the implications of design choices). In total, this dataset contains $295$ questions, averaging $59$ questions per video.

\begin{table}
\caption{Summary statistics for our two datasets: Ciena and Course.}
\label{tab:datasets}
\vspace*{-.2cm}
\begin{center}
\scalebox{0.75}{
\begin{tabular}{p{7.5cm}| p{1.5cm} |p{1.2cm}}
\toprule
  & \textbf{Ciena} & \textbf{Course}\\
  \hline
\# of videos&5&5\\
\hline
Average video duration &55.07min&72.60min\\
\hline
Average video size&271.2MB&97.72MB\\
\hline
Average \# of chunks per video&679.8&1003.8\\
\hline
Average \# of diagram-related keyframes per video&14.8&70.4\\
\hline
Average \# of non-diagram related keyframes per video&48&72.2\\
\hline
Average \# of diagram-related questions per video&4.2 &28.2\\
\hline
Average \# of visual-content-related questions per video &5.8 &15.4\\
\hline
Average \# of audio-related questions per video& 37.2 &15.4\\
\hline
Total \# of questions& 236 & 295\\
\bottomrule
\end{tabular}}
\end{center}
\vspace*{-.6cm}
\end{table}

\subsection{Experiment Setting}

\begin{table}[t]
\centering
\caption{Main parameters and LLM models of \name.}\label{tab:exp_params}\vspace*{-.1cm}
\scalebox{0.95}{\begin{tabularx}{\columnwidth}{X l}
\toprule
 \textbf{Parameter/Model} & \textbf{Value} \\
\midrule
 Sampling rate                         & 1 FPS \\
 SSIM threshold                        & 0.92 \\
 Top candidate chunks $K$              & 200 \\
 Keyframe Classifier            & GPT-4o-mini \\
 Description Extractor (non-diagram-related frames)    & GPT-4o \\
 Description Extractor (diagram-related frames)    & GPT-5 \\
 Diagram Category Detection & GPT-4o-mini\\
QA model                              & GPT-4o \\
Temperature                  & 0 \\
\bottomrule
\end{tabularx}}

\vspace{0.35em}
\begin{minipage}{0.92\columnwidth}
{\scriptsize
\emph{Note:} The LLM roles specified in Fig.~\ref{fig:overview} correspond to the following actual LLMs: lightweight LLM = GPT-4o-mini, LLM = GPT-4o, and frontier LLM = GPT-5.}
\end{minipage}
\vspace*{-.45cm}
\end{table}

Table~\ref{tab:exp_params} shows the configuration parameters used for \name\  in our experiments. Based on preliminary experiments, we set the sampling rate to $1$ FPS, so that only one frame per second is retained, which substantially reduces computational cost while preserving sufficient temporal resolution for reasoning. We set the SSIM threshold to $0.92$ for keyframe selection, following the SSIM-based redundancy filtering described in Section~\ref{sec:video-preprocessing}. With this threshold, we discard only frames that are nearly identical while retaining those with even subtle differences. This is important for technical meeting videos, where small visual updates, such as new text, arrows, or diagram elements appearing on a slide, introduce important information that should be preserved as distinct frames. For answer generation, we start by keeping the top $200$ most relevant chunks, chosen to balance evidence coverage and prompt size. If the retrieved context exceeds GPT-4o’s context-length limit, we retain only the highest-ranked chunks that fit within the model’s context window and omit the remaining lower-ranked chunks, thereby minimizing information loss.

In the visual description extraction, we use GPT-4o-mini to classify keyframes and detect diagram categories, GPT-4o to extract descriptions from non-diagram related frames, and GPT-5 to generate descriptions for  diagram-related frames. We also use GPT-4o for answer generation. All OpenAI models used in our experiments, including GPT-5, were accessed through Ciena-approved OpenAI API subscriptions at the time of the study. For all LLM queries, we set the temperature to zero to ensure that the outputs are as deterministic and reproducible as possible.

To our knowledge, no prior work has addressed open-ended question answering over technical videos. For comparison with existing methods, we use DrVideo~\cite{Ma:25} as the closest and most recent approach to our work. DrVideo is a state-of-the-art retrieval-based method for long-video question answering. However, it is designed for multiple-choice questions, whereas all questions in our datasets are open-ended. We therefore adapt DrVideo by extending its prompts and post-processing pipeline to generate open-ended responses in a compact JSON format with \texttt{final\_answer} and \texttt{rationale}; our adapted DrVideo baseline is released as part of our replication package~\cite{loviqa-prompts}. We use GPT-4o as the underlying LLM for DrVideo, consistent with the LLM used for \name.

We used the same experimental setup for both \name\ and DrVideo. For both methods, all local preprocessing, retrieval, and logging steps were executed on a machine equipped with an Intel(R) Core(TM) Ultra 7 165U CPU and $16$ GB of memory. Audio transcription was performed locally using Whisper~\cite{radford2023whisper}. All LLM calls for both \name\ and DrVideo, including calls to GPT-4o-mini, GPT-4o, and GPT-5, were made through Ciena-approved OpenAI API subscriptions rather than locally deployed models. This setup ensured that \name\ and DrVideo were evaluated under the same execution environment and API access conditions. All code and experimental results are available online~\cite{loviqa-prompts}.

\subsection{Metrics}
\label{subsec:metric}
To answer RQ1, we evaluate the accuracy of answers generated by \name\ and the baseline by comparing them against the ground-truth answers. The evaluation was conducted by two labelers who were not authors of this study and were not involved in constructing the datasets. Both labelers received detailed annotation guidelines and assessed each generated answer against its corresponding ground truth. Table~\ref{tab:prediction-examples} presents representative examples of questions, generated answers, ground-truth answers, and assigned labels, while Table~\ref{tab:criteria} defines the criteria for judging semantic equivalence. For each pair of  generated and ground-truth answers, the labelers assigned a binary label: \emph{correct} or \emph{incorrect}. An answer was labeled incorrect if it contradicted the ground truth, failed to answer the question, omitted required information, or contained incorrect numerical values, units, or entities. As shown in Table~\ref{tab:prediction-examples}, the first two generated answers were labeled correct because they satisfy the criteria in Table~\ref{tab:criteria}; the third is incorrect because it violates criteria (ii) and (iii).

\begin{table}[t]
\caption{Examples of questions, generated answers, corresponding ground-truth answers, and correctness of the generated answers. Redacted text appears in square brackets([]) and is highlighted in \textcolor{red}{red}.}
\label{tab:prediction-examples}
\begin{center}
\vspace*{-.2cm}
\scalebox{0.85}{\begin{tabularx}{0.55\textwidth}{l X}
\hline
\textbf{Item} & \textbf{Content} \\
\hline

\textbf{Example 1} & \textbf{Correct according to the criteria in  Table~\ref{tab:criteria}} \\

Question &
Who are the main vertically integrated vendors in the coherent plug market? \\

Generated Answer &
\textcolor{red}{[Vendor1]}, \textcolor{red}{[Vendor2]}, \textcolor{red}{[Vendor3]}, and \textcolor{red}{[Vendor4]}. \\

Ground truth &
\textcolor{red}{[Vendor1]}, \textcolor{red}{[Vendor2]}, \textcolor{red}{[Vendor3]}, and \textcolor{red}{[Vendor4]} are the main vertically integrated vendors producing plugs and having both DSP and electro-optics capabilities. \\

\hline

\textbf{Example 2} & \textbf{Correct according to the criteria in Table~\ref{tab:criteria}} \\

Question &
How much total spectrum is required to transport 2~Terabits (2T) of data using \textcolor{red}{[Internal Transponder A]} versus \textcolor{red}{[Internal Transponder B]}, as shown in the table and diagrams? \\

Generated Answer &
\textcolor{red}{[Internal Transponder A]}: 625~GHz; \textcolor{red}{[Internal Transponder B]}: 400~GHz. \\

Ground truth &
To transport 2T of data, \textcolor{red}{[Internal Transponder A]} requires 625~GHz of spectrum, whereas \textcolor{red}{[Internal Transponder B]} requires only 400~GHz of spectrum. \\

\hline

\textbf{Example 3} & \textbf{Incorrect as it violates the criteria ii and iii in Table~\ref{tab:criteria}} \\

Question &
On the \textit{``STATES''} slide fragment, which example state is shown receiving two incoming \textit{``buttonOrObstacle''} arrows? \\

Generated Answer &
Opening \\

Ground truth &
HalfOpen \\

\hline
\end{tabularx}}
\end{center}
\vspace*{-.6cm}
\end{table}

\begin{table}[t]
\caption{Correctness criteria for generated answers.}
\label{tab:criteria}
\begin{center}
\vspace*{-.3cm}
\scalebox{0.9}{
\begin{tabular}{|>{\raggedright\arraybackslash}m{3.3cm}|>{\raggedright\arraybackslash}m{5.5cm}|}
\hline
\textbf{Criterion} & \textbf{Description} \\
\hline

\textbf{i. Exact numeric agreement (when applicable)}
& For questions requiring quantitative answers, the response must exactly match the numeric values and units in the ground truth, or present an equivalent representation that unambiguously conveys the same quantity. Differences in formatting are allowed, but any discrepancy in values, units, or referenced entities renders the answer incorrect. \\
\hline

\textbf{ii. Complete coverage of question-required content}
& For questions with ground-truth answers that list multiple entities or facts (e.g., vendors or required conditions), the generated answer must contain all listed items. Missing any item is considered incorrect. \\
\hline

\textbf{iii. Core-answer equivalence under extended ground truth}
& If the ground-truth answer includes extra explanatory details, the generated answer is still correct as long as it conveys the essential facts or claims that directly answer the question and does not contradict the ground truth. \\
\hline

\end{tabular}}
\end{center}
\vspace*{-.6cm}
\end{table}

\begin{table}[t]
\caption{Inter-rater agreement (Cohen’s $\kappa$) per split and pooled.}\label{tab:kappa}
\centering
\vspace*{-.2cm}
\small
\scalebox{0.8}{\begin{tabular}{p{1.6cm}|p{1.5cm}|p{2cm}|p{1cm}}
\toprule
\textbf{Split}  & \textbf{Cohen's $\kappa$}&\textbf{Interpretation}&\textbf{dataset} \\
\midrule
\multirow{2}{*}{DrVideo}   & 0.750&Substantial&Ciena \\
\cline{2-4}   & 0.929&Almost perfect&Course \\
\hline
\multirow{2}{*}{\name}   & 0.621&Substantial&Ciena \\
\cline{2-4}  & 0.967&Almost perfect&Course \\
\midrule
\textit{Pooled (all)}  & 0.908&Almost perfect &\multicolumn{1}{c}{---}\\ 
\bottomrule
\end{tabular}}
\vspace*{-.45cm}
\end{table}

Interrater agreement among the two labelers was measured using Cohen's kappa ($\kappa$). As shown in Table~\ref{tab:kappa},  agreement between the two labelers is consistently high. For DrVideo, we obtain $\kappa=0.750$ on the Ciena dataset and $0.929$ on the Course dataset, corresponding to  substantial  and  almost perfect agreement respectively. For \name, $\kappa=0.621$ on the Ciena dataset (substantial agreement) and $0.967$ on the Course dataset (almost perfect agreement). The pooled agreement across all items is $\kappa=0.908$, indicating near-perfect reliability of the binary judgments~\cite{landis1977measurement}. A disagreement was counted whenever the two labelers assigned different labels to the same answer.  All disagreements were subsequently resolved in a consensus meeting, during which the labelers discussed each disputed case and jointly agreed on the final label. Accuracy was then calculated as the proportion of questions labeled as correct out of the total number of questions.

For RQ2, we evaluate efficiency using three metrics: \emph{offline video-indexing time}, \emph{interactive answering time per question}, and \emph{LLM API cost}. We distinguish offline and interactive time to avoid conflating one-time preprocessing with the latency experienced by users during question answering. \name\ has a two-stage design: it first processes each video once to construct a reusable, time-stamped multimodal evidence corpus, and then uses this corpus to answer user questions through retrieval and generation. We define the three metrics measuring efficiency as follows:  (1)~Offline video-indexing time measures the wall-clock time needed to process a video before it becomes queryable, including audio and visual extraction, description generation, chunking, embedding, and vector-database storage. This metric captures the one-time preprocessing cost that \name\ shifts offline and reuses across questions. (2)~Interactive answering time per question measures the user-facing latency from question submission to answer generation after indexing, including query embedding, retrieval, prompt construction, and answer generation. (3)~LLM API cost measures the token-based cost of commercial LLM calls. We report the average cost per video, including indexing and answer generation for all questions associated with that video.

For RQ3, we collected qualitative feedback from three Ciena engineers through hands-on use of \name\ followed by semi-structured interviews. At the start of each session, we demonstrated \name\ using a randomly selected internal Ciena technical meeting video and briefly explained the topic and scope of the assigned video to provide participants with sufficient context for formulating relevant questions. Participants then completed approximately ten minutes of guided practice to become familiar with the interface and the question-answering workflow. After this guided practice, participants were given access to the interactive interface of \name\ and were free to explore the assigned video in a hands-on manner by watching or skimming relevant segments, navigating to timestamps, and asking natural-language questions about both its spoken and visual content. Participants continued using \name\ until they felt they had sufficiently explored its capabilities, after which they completed the questionnaire described below.

The questionnaire consisted of five parts. The first part captured participants’ experience with company technical meeting videos, including their viewing duration and frequency. The second part examined their purposes for watching technical meeting videos and the challenges they encounter when consuming them. The third part asked about any alternative tools they had previously used for similar tasks. The fourth part elicited their views on the usefulness, accuracy, and effectiveness of \name. Finally, the fifth part invited participants to suggest  future improvements. The completed questionnaires for all three participants are available online~\cite{loviqa-prompts}.

\subsection{Results}

\noindent\textbf{RQ1: Accuracy.} To answer RQ1, we assess the accuracy of both \name\ and the baseline,  DrVideo, using the accuracy metric described in Section~\ref{subsec:metric}.   Table~\ref{tab:rq1} summarizes the accuracy results achieved by \name\ and DrVideo on our two datasets.  Specifically, for the Ciena dataset, the accuracy (i.e., the rate of correct answers) improves from $31\%$ (DrVideo) to $94\%$ (\name), representing an absolute gain of $63\%$ and a relative improvement of $203\%$. For the Course dataset, the accuracy increases from $21\%$ to $88\%$, corresponding to an absolute gain of $67\%$ and a relative improvement of $319\%$.

\begin{table}
\caption{Accuracy comparison between \name\ (denoted by L) and DrVideo (denoted by D).}
\label{tab:rq1}
\vspace*{-.2cm}
\begin{center}
\scalebox{0.9}{
\begin{tabular}{>{\centering\arraybackslash}m{0.9cm}|>{\centering\arraybackslash}m{1.8cm}|>{\centering\arraybackslash}m{1.8cm}|>{\centering\arraybackslash}m{1.2cm}}
\toprule
& \textbf{\# of correct answers(L-D)}
& \textbf{\# of incorrect answers(L-D)}
& \textbf{Accuracy (L-D)} \\
\hline
\textbf{Ciena} & 222-74 & 14-162 & 0.94-0.31 \\
\hline
\textbf{Course} & 261-63 & 34-232 & 0.88-0.21 \\
\bottomrule
\end{tabular}}
\end{center}
\vspace*{-.45cm}
\end{table}

As shown in Table~\ref{tab:datasets}, the Course dataset contains more diagram-related content than the Ciena dataset. On average, each Course video contains $70.4$ diagram-related keyframes, compared with $14.8$ in the Ciena dataset. The Course dataset also contains more diagram-related questions per video ($28.2$ vs. $4.2$). The improvement achieved by \name\ is more pronounced for the Course dataset than for the Ciena dataset ($319\%$ vs. $203\%$ in accuracy). While other factors may contribute to this gap, these results suggest that \name\ may be particularly useful for diagram-rich technical videos.\\[-.2cm]

\resq{The answer to \textbf{RQ1} is that \name\ achieves substantially higher accuracy than a state-of-the-art approach, DrVideo, improving accuracy from $31\%$ to $94\%$ on the Ciena dataset and from $21\%$ to $88\%$ on the Course dataset.} 

\noindent\textbf{RQ2: Efficiency.} We evaluate efficiency using the three metrics defined in Section~\ref{subsec:metric}: \emph{offline video-indexing time}, \emph{interactive answering time per question}, and \emph{LLM API cost}. We report the cost per video, including both the one-time indexing cost and the answer-generation cost for all questions associated with that video. Recall that \emph{offline video-indexing time} measures the one-time preprocessing cost required for \name\ to construct a reusable multimodal evidence corpus for each video, including audio transcription, visual description extraction, chunk generation, embedding, and vector database construction. In contrast, \emph{interactive answering time per question} measures the latency experienced by users during question answering once indexing has been completed. 

Table~\ref{tab:rq2} reports the average values and standard deviations for all three metrics across both datasets. Since DrVideo does not construct a reusable offline index and instead repeatedly processes raw video content during question answering, the offline video-indexing metric is not applicable to the baseline and is therefore marked as N/A.

\begin{table}[t]
\caption{Efficiency comparison between \name\ and DrVideo on the Ciena and Course datasets. Results show mean (std.) offline indexing time, per-question latency, and LLM API cost.}
\label{tab:rq2}
\vspace*{-.1cm}
\centering
\scriptsize
\setlength{\tabcolsep}{2.3pt}
\renewcommand{\arraystretch}{1.18}
\begin{tabular}{@{}p{0.95cm}|p{1.85cm}|p{1.95cm}|p{2.35cm}|p{0.85cm}@{}}
\toprule
 & 
\makecell[l]{\textbf{offline}\\\textbf{video-indexing}\\\textbf{time}\\\textbf{avg (std)}} &
\makecell[l]{\textbf{interactive}\\\textbf{answering time}\\\textbf{per question}\\\textbf{avg (std)}} &
\makecell[l]{\textbf{LLM API cost}\\\textbf{per video}\\\textbf{avg (std)}} &
\textbf{dataset} \\
\midrule
\multirow{2}{*}{\textbf{DrVideo}}
& N/A & $81.3$s ($24.8$s) & $149$ USD ($29.3$ USD) & Ciena \\
& N/A & $98.4$s ($25.0$s) & $156$ USD ($59.1$ USD) & Course \\
\midrule
\multirow{2}{*}{\textbf{\name}}
& $1.47$h ($0.31$h) & $3.3$s ($1.2$s) & $36$ USD ($3.5$ USD) & Ciena \\
& $2.71$h ($1.18$h) & $9.2$s ($7.0$s) & $38$ USD ($12.7$ USD) & Course \\
\bottomrule
\end{tabular}
\vspace*{-.8cm}
\end{table}

For \name, the average offline video-indexing time is $1.47$ hours per video on the Ciena dataset and $2.71$ hours per video on the Course dataset. Although this preprocessing step incurs an upfront cost, it is performed only once per video, can be executed offline (e.g., overnight), and enables efficient reuse across subsequent queries.

The benefits of this design are reflected in the interactive answering latency. On the Ciena dataset, DrVideo requires an average of $81.3$ seconds per question, whereas \name\ answers questions in only $3.3$ seconds after indexing, yielding approximately a $24.6$ times  speedup. Similarly, on the Course dataset, DrVideo requires $98.4$ seconds per question, while \name\ reduces this latency to $9.2$ seconds, corresponding to an approximate $10.7$ times improvement. These results indicate that shifting computation to an offline indexing stage substantially improves responsiveness during user interaction.

\name\ also achieves substantially lower LLM API costs than DrVideo. On the Ciena dataset, \name\ incurs an average cost of $36$ USD per video, compared with $149$ USD for DrVideo. On the Course dataset, \name\ costs $38$ USD per video, whereas DrVideo costs $156$ USD. Overall, this corresponds to cost reductions of approximately $76\%$ and $75\%$ on the Ciena and Course datasets, respectively. Since most of \name's cost is incurred during one-time offline indexing, the marginal LLM API cost of each subsequent query is low, making the approach practical when videos are queried multiple times.

\resq{The answer to \textbf{RQ2} is that \name\ improves interactive efficiency by shifting expensive video processing to a one-time offline indexing stage. After indexing, \name\ answers questions substantially faster than DrVideo, reducing average per-question latency from $81.3$s to $3.3$s on Ciena and from $98.4$s to $9.2$s on Course. It also lowers LLM API cost by approximately $75\%$ across both datasets.}

\noindent\textbf{RQ3: Practitioner Feedback.} To address RQ3, we collected qualitative feedback from three Ciena engineers through hands-on use of \name\ followed by semi-structured interviews, as described in Section~\ref{subsec:metric}. The completed questionnaires from all three participants are available online~\cite{loviqa-prompts}. Here, we outline the main findings.

All participants had at least three years of experience at Ciena. They regularly use technical meeting recordings to  synchronize with ongoing projects, learn about new technologies and system introductions, understand the rationale behind technical decisions, and identify stakeholder requirements and constraints. None of the participants is a co-author of this paper. The interviews highlighted some recurring challenges associated with reviewing long technical meeting videos. Participants noted that the length of the recordings makes them time-consuming to revisit and makes it difficult to remember specific details afterward. In practice, engineers often revisit recordings with focused information needs, such as clarifying a design decision or locating a particular requirement, and therefore prefer not to watch the entire video again to find relevant information.

In addition, participants  discussed the tools they currently use to support video review. One participant reported using Zoom’s meeting assistant, while another relied on several general-purpose generative AI tools, including GPT, Claude, and YouTube’s AI-assisted video features. However, participants emphasized several limitations of these tools when applied to long technical meeting videos. First, directly uploading long videos to general-purpose LLM systems often leads to unpredictable processing times and, in some cases, unusable or incomplete answers. Second, these systems are not specifically designed for software-engineering artifacts commonly found in technical meetings, such as UML diagrams, architecture diagrams, workflow diagrams, and other structured engineering visualizations. As a result, they may fail to correctly interpret diagram semantics or connect visual evidence with spoken explanations. Finally, participants noted that these tools typically do not provide explicit time-stamped multimodal evidence tracing, making it difficult to verify answers or navigate back to the relevant video segments.

Two participants reported that the answers generated by \name\ during the evaluation sessions were accurate and easy to understand. The third participant asked three questions and indicated that \name\ answered two correctly, while one answer was only partially accurate because it required identifying a very small component within a diagram. Despite this limitation, all participants stated that \name\ successfully traced answers back to the relevant portions of the video. Even in the partially correct case, the participant considered the identified video segment helpful because it substantially narrowed the search space for manual verification.

Overall, two participants rated \name\ as accurate or very accurate. More importantly, all three participants emphasized that the ability to locate and trace relevant evidence within long technical meeting recordings was one of the most valuable aspects of the system. These results suggest that, beyond answer generation alone, \name\ provides practical value through grounded multimodal retrieval and time-stamped source tracing for diagram-rich technical meeting videos.

\resq{The answer to RQ3 is that, based on practitioner feedback from three Ciena engineers, \name\ was perceived as useful for locating, verifying, and interpreting information in long technical meeting videos. Two participants rated \name\ as accurate or very accurate, and all three found its time-stamped source tracing helpful for verifying answers and locating relevant video segments.}

\subsection{Validity Considerations and Limitations}
\label{subsec:validity}

\noindent\textbf{\emph{Internal Validity.}}  A potential threat to internal validity arises from the inherent randomness of LLMs. Due to the high execution time, we were unable to run \name\ or the baseline multiple times to account for randomness. To mitigate this threat, we set the LLM temperature parameter to zero to improve determinism in the generated outputs. In addition, the substantial number of questions in our datasets -- averaging $47.2$ questions per video for the Ciena dataset and $59$ questions per video for the Course dataset -- helps compensate for the lack of repeated runs.

\noindent\textbf{\emph{External Validity.}} 
Although our evaluation uses two datasets spanning both an industrial setting (Ciena) and a public course setting, the total number of videos remains limited to ten videos, which makes the study appropriate as an initial assessment rather than a definitive demonstration of generalizability. The videos cover different presentation styles, question types, and varying amounts of diagram-centric content, but differences in domains and  meeting practices may still affect how well the results transfer to other real-world contexts. Furthermore, \name\ relies on OpenAI LLMs -- GPT-4o-mini for keyframe classification and diagram category detection, GPT-4o for non-diagram description extraction and answer generation, and GPT-5 for diagram-related frame processing -- so the reported outcomes may vary with other LLMs due to differences in their capabilities and behavior.

 \section{Related Work}
\label{subsec:related}
Table~\ref{tab:related-work} positions \name\ within the literature by comparing it to relevant approaches across three capabilities central to technical meeting question answering (QA): \textbf{(i)~long-video support}, i.e., the ability to process hour-scale videos; \textbf{(ii)~audio coverage}, i.e., whether spoken content is used as evidence; and \textbf{(iii)~diagram grounding}, i.e., whether the method explicitly supports \hbox{understanding and reasoning over diagrams.}

\begin{table}[t] \caption{Related work organized by three capabilities central to video question-answering: long-video support, audio coverage, and diagram grounding. In this table, \cmark\ indicates full support, \xmark\ indicates no support, and \pmark\ indicates partial support.}\label{tab:related-work} 
\vspace*{-0.4cm}
\begin{center} 
\scalebox{0.75}{\begin{tabular}{|p{4.0cm}|c|c|c|} 
\hline
\textbf{Approach(es)} & \textbf{\textit{Long-video}} & \textbf{\textit{Audio}} & \textbf{\textit{Diagram}} \\ \hline \makecell[l]{CodeTube~\cite{Ponzanelli:ICSE16CodeTube},\\ Escobar-Avila et al.~\cite{EscobarAvila:ICSEC17VideoTagging},\\ VT-Revolution~\cite{Bao:TSE19VTRevolution},\\ psc2code~\cite{Bao:TOSEM20psc2code},\\ PSFinder~\cite{Yang:22},\\ LongVLM~\cite{longvlm},\\ Video-LLaVA~\cite{video-llava}} & \cmark & \xmark & \xmark \\ \hline \makecell[l]{Flamingo~\cite{flamingo},\\ Video-ChatGPT~\cite{video-chatgpt}} & \xmark & \xmark & \xmark \\ \hline Video-LLaMA~\cite{video-llama} & \xmark & \cmark & \xmark \\ \hline \makecell[l]{PlotQA~\cite{plotqa}, ChartQA~\cite{chartqa},\\ ChartVQA~\cite{chartvqa}, AI2D~\cite{ai2d},\\ DocVQA~\cite{docvqa}} & \xmark & \xmark & \cmark \\ \hline DrVideo~\cite{Ma:25} & \cmark & \pmark & \pmark \\ \hline \textbf{\name} & \textbf{\cmark} & \textbf{\cmark} & \textbf{\cmark} \\ \hline \end{tabular}} \end{center} 
\vspace*{-.5cm}
\end{table}

A first line of work in software-engineering video analysis focuses on retrieval, navigation, and code extraction rather than question answering. CodeTube~\cite{Ponzanelli:ICSE16CodeTube} and the approach of Escobar-Avila et al.~\cite{EscobarAvila:ICSEC17VideoTagging} identify relevant segments in software tutorial videos using tag-based retrieval. VT-Revolution~\cite{Bao:TSE19VTRevolution} and psc2code~\cite{Bao:TOSEM20psc2code} extend this direction by supporting interactive navigation and code extraction from programming videos. PSFinder~\cite{Yang:22} similarly enables efficient search over live-coding screencasts. These methods are relevant because they operate over long videos in software-related settings; however, they are not designed for grounded, open-ended question answering, do not integrate audio evidence, and do not explicitly reason over diagrams  in slides or screen shares.

A second line of work studies general-purpose multimodal video-language models. Flamingo~\cite{flamingo} and Video-ChatGPT~\cite{video-chatgpt} extend image-language modeling to video through sparse frame sampling and prompt-based generation. These models are effective for clip-level captioning and question answering, but they are not intended for hours-long videos and rely primarily on visual input. Video-LLaVA~\cite{video-llava} moves further toward video understanding by aligning frame sequences with instruction tuning, and LongVLM~\cite{longvlm} introduces memory- and retrieval-based mechanisms for long-video processing. While these approaches improve temporal coverage, they still do not explicitly incorporate audio as a first-class source of evidence and do not provide dedicated support for diagram reasoning. Video-LLaMA~\cite{video-llama} addresses part of this limitation by incorporating audio into an instruction-tuned audio-visual LLM. However, it still focuses on short- to medium-length videos rather than hour-long technical recordings with diagram-heavy content.

A third, complementary line of work comes from diagram and document question answering. PlotQA~\cite{plotqa}, ChartQA~\cite{chartqa}, and ChartVQA~\cite{chartvqa} study reasoning over charts and plots, including reading labels, inferring values, and performing logical or arithmetic operations. AI2D~\cite{ai2d} expands the scope to general diagram understanding by modeling diagram elements and their relations, while DocVQA~\cite{docvqa} focuses on question answering over document images. These studies are highly relevant to our setting because they highlight the importance of structured visual reasoning. However, they are fundamentally image-centric: they do not handle long temporal context, do not integrate spoken explanations, and therefore do not directly address technical meeting videos where answers often depend on combining diagram content with narration over time.

The closest prior work to ours is DrVideo~\cite{Ma:25}, which we use as our baseline. DrVideo is a retrieval-based framework for long-video understanding that retrieves question-conditioned evidence and iteratively refines it into a document for answering, making it more suitable for hour-scale videos than clip-oriented video QA systems. However, DrVideo only partially addresses technical meeting QA. It lacks explicit mechanisms for detecting diagram-relevant segments, categorizing diagram types, or adapting answer generation to diagram structure. As a result, it can answer diagram-related questions only when the retrieved textual evidence captures the necessary visual semantics. Moreover, DrVideo is designed for multiple-choice QA, so applying it to open-ended questions requires additional prompting and post-processing. Overall, no prior approach is tailored to technical meeting video QA, which requires long-video processing, audio-visual evidence integration, and grounding in structured artifacts such as software diagrams. Existing methods address these needs only partially. \name\ fills this gap by constructing a reusable audio-visual evidence corpus, explicitly detecting and categorizing diagrams, and using diagram-aware prompting for structured visual evidence.

 \section{Lessons Learned}
\label{sec:lessons}

\textbf{\emph{Lesson 1: Customization for software-engineering diagram semantics is important for diagram-rich technical videos.}} Technical engineering meetings often convey essential information through diagrams that capture structure, dependencies, relationships, and design intent. Our results suggest that treating these diagrams as generic visual content can limit answer accuracy. \name\ is customized for software-engineering artifacts by detecting diagram-rich frames, classifying diagram types, and extracting diagram-specific semantics before generating answers from visual and audio evidence. To evaluate the importance of this capability, we compared \name\ against DrVideo~\cite{Ma:25}, a recent state-of-the-art long-video question-answering approach from the literature, which we adapted to support open-ended questions but which does not explicitly support software-engineering diagram semantics. Our RQ1 results provide evidence for the value of this customization. \name\ improves accuracy from 31\% to 94\% on the Ciena dataset and from 21\% to 88\% on the Course dataset. While other factors may contribute to these gains, the results suggest that semantic-level understanding of software-engineering diagrams is important for effective question answering over diagram-rich technical meeting videos.

\textbf{\emph{Lesson 2: Brute-force frame analysis is costly and unnecessary.}} RQ2 shows that \name\ is substantially more efficient and less costly than DrVideo, a state-of-the-art baseline. A key reason is that \name\ samples frames at fixed intervals (1 FPS in our implementation), removes visually redundant sampled frames using SSIM, and applies LLM-based visual description extraction only to the retained keyframes. This reduces repeated analysis of visually similar frames while preserving the key visual evidence needed for question answering.  As shown in our evaluation, this selective frame analysis reduces unnecessary LLM calls without sacrificing answer quality. As a result, \name\ lowers per-question latency from 81.3s to 3.3s on Ciena and from 98.4s to 9.2s on Course, while reducing LLM API cost by approximately 75\%. These results show that vision-based similarity metrics~\cite{delaram} can effectively identify the focused visual evidence needed for technical video question answering, rather than processing all frames exhaustively.

\textbf{\emph{Lesson 3: Grounded evidence matters as much as answer generation.}} A third lesson is that \name’s value lies not only in answering questions, but also in helping users navigate long technical videos through grounded multimodal evidence. Participants noted that such videos are long, easy to forget, and hard to search afterward. In this setting, \name’s ability to connect answers to relevant video segments was a key differentiating factor: all three participants rated its source tracing as helpful. One participant noted that \name\ helped them “jump to the sections I need”, while others emphasized that it enabled them to quickly find specific information without rewatching the entire video. Participants also valued its diagram-related support, with two rating diagram-related answers as accurate or very accurate and one reporting that it made diagram understanding very easy. Overall, these comments suggest that \name’s distinctive strength is the combination of diagram analysis, question answering, and traceable evidence, rather than answer generation alone. At the same time, the feedback shows that fine-grained diagram understanding and the user experience of interacting with long videos remain important directions for future research.

\textbf{\emph{Deployment at Ciena.}} \name\ was developed in collaboration with Ciena over seven months and deployed as part of Ciena's AI-based solutions. It uses Milvus~\cite{Wang:SIGMOD21Milvus} as a Ciena-hosted vector database for reusable offline indexing. For each archived meeting video, \name\ extracts audio and visual evidence, builds time-stamped multimodal chunks, embeds them, and stores them in Milvus for efficient retrieval. Engineers can then query the pre-indexed archive without reprocessing videos for each question.

 \section{Conclusion}
In this paper, we presented \name, an LLM-based multimodal question-answering system for diagram-rich technical meeting videos. Across an industrial dataset and a public-domain dataset, \name\ improves average accuracy over a state-of-the-art baseline, DrVideo, from $31\%$ to $94\%$ and from $21\%$ to $88\%$, respectively, while reducing average LLM API cost by about $75\%$. These results show that diagram-aware processing, reusable video indexing, and evidence-grounded answer generation are key to practical QA over long technical recordings. More broadly, \name\ can help engineers recover rationale, revisit historical decisions, and understand architecture and requirements discussions. Feedback from Ciena engineers suggests that future work should further improve video navigation, answer verification, and user interaction.%
 
\section{Appendix}
\label{appendix}

\subsection{Prompt Outlines}
\label{appendix:prompt-outlines}

\begingroup
\setlength{\fboxsep}{3.2pt}
\setlength{\fboxrule}{0.35pt}

\newcommand{\LmvqaPromptBox}[2]{\noindent\fbox{\begin{minipage}{0.985\linewidth}
\raggedright\footnotesize
\textbf{#1}\par\vspace{0.08em}
#2
\end{minipage}}\par\vspace{0.25em}
}

\newcommand{\LmvqaPItem}[2]{\noindent\textbf{#1} #2\par
}

\vspace{0em}

\LmvqaPromptBox{(Listing a) Prompt Outline for Diagram-Like Frame Detection}{
\LmvqaPItem{(I)}{Role: a fast triage classifier.}
\LmvqaPItem{(II)}{Input: a single video frame image.}
\LmvqaPItem{(III)}{Task: decide whether the frame is diagram-like. Diagram-like visuals include diagrams, flowcharts, schematics, architecture diagrams, charts, tables, UI wireframes, whiteboards with boxes or arrows, and equations.}
\LmvqaPItem{(IV)}{Output: \texttt{LABEL: <DIAGRAM|NOT\_DIAGRAM>}; \texttt{TYPES: <types>}; \texttt{CONF: <0.00-1.00>}.}
}

\LmvqaPromptBox{(Listing b) Prompt Outline for Non-Diagram Frame Captioning}{
\LmvqaPItem{(I)}{Role: an assistant analyzing a video frame from an educational lecture or tutorial.}
\LmvqaPItem{(II)}{Frame metadata: \texttt{\{timestamp\}}.}
\LmvqaPItem{(III)}{Input: a sampled video frame image.}
\LmvqaPItem{(IV)}{Task: describe visible slide, screen, chart, or non-diagram content; extract and summarize visible text when useful.}
\LmvqaPItem{(V)}{Grounding rules: ignore irrelevant UI elements, avoid redundancy, and describe only visible content.}
\LmvqaPItem{(VI)}{Output: a fluent structured English frame description.}
}

\LmvqaPromptBox{(Listing c) Prompt Outline for Diagram Classification}{
\LmvqaPItem{(I)}{Role: a fast diagram-type classifier.}
\LmvqaPItem{(II)}{Input condition: the image is already known to be diagram-like.}
\LmvqaPItem{(III)}{Categories: \texttt{UML\_CLASS}, \texttt{UML\_SEQUENCE}, \texttt{UML\_STATE\_ACTIVITY}, \texttt{UML\_USE\_CASE}, \texttt{NETWORK\_TOPOLOGY}, \texttt{ARCH\_WORKFLOW}, or \texttt{OTHER}.}
\LmvqaPItem{(IV)}{Output: \texttt{CATEGORY: <category>}; \texttt{CONF: <0.00-1.00>}.}
}

\LmvqaPromptBox{(Listing d) Definitions for diagram categories used in prompt outline in listing c. 
}{
\LmvqaPItem{\texttt{UML\_CLASS}:}{diagram title; scope or packages; classes and interfaces; relationships; key takeaways.}
\LmvqaPItem{\texttt{UML\_SEQUENCE}:}{diagram title; participants or lifelines; temporally ordered messages; activations; combined fragments; key scenario summary.}
\LmvqaPItem{\texttt{UML\_STATE\_ACTIVITY}:}{diagram title; subtype; nodes; transitions or flows; composite states, swimlanes, forks, joins, and behavior summary.}
\LmvqaPItem{\texttt{UML\_USE\_CASE}:}{diagram title; system boundary; actors; use cases; associations; include, extend, and generalization relationships.}
\LmvqaPItem{\texttt{NETWORK\_TOPOLOGY}:}{title or legend; nodes and endpoints; links; traffic or flows; key topology takeaway.}
\LmvqaPItem{\texttt{ARCH\_WORKFLOW}:}{title; components or modules; interfaces or artifacts; connections or flows; workflow steps.}
\LmvqaPItem{\texttt{OTHER}:}{title; diagram type guess; main elements; relationships or structure; key takeaway.}
}

\LmvqaPromptBox{(Listing e) Prompt Outline for Diagram Frame Extraction}{
\LmvqaPItem{(I)}{Role: a diagram-frame describer specialized according to the selected diagram category.}
\LmvqaPItem{(II)}{Frame metadata: \texttt{\{timestamp\}}.}
\LmvqaPItem{(III)}{Input: the diagram-like video frame image.}
\LmvqaPItem{(IV)}{Generic schema: title, purpose, global layout, nodes, edges, flow sequence, text and labels, data or variables, context anchors, and key takeaway.}
\LmvqaPItem{(V)}{Category-specific schema: refined according to the diagram classifier output.}
\LmvqaPItem{(VI)}{Grounding rules: describe only visible content, quote labels when possible, mark unclear text as \texttt{[illegible]}, and do not invent facts.}
\LmvqaPItem{(VII)}{Output: exhaustive structured plain English text, not JSON.}
}

\LmvqaPromptBox{(Listing f) Prompt Outline for RAG-Based Video Question Answering}{
\LmvqaPItem{(I)}{Role: an assistant answering questions about technical videos using time-stamped multimodal information.}
\LmvqaPItem{(II)}{Video metadata: \texttt{\{video\_duration\}}.}
\LmvqaPItem{(III)}{Retrieved context: \texttt{\{retrieved\_context\}}, formatted as \texttt{[source] <timestamp>: <text>}.}
\LmvqaPItem{(IV)}{User question: \texttt{\{user\_question\}}.}
\LmvqaPItem{(V)}{Grounding rules: answer only from retrieved segments; do not introduce unsupported information; use the insufficient-information answer when needed.}
\LmvqaPItem{(VI)}{Output: \texttt{Answer:} plus final answer, and \texttt{Used\_context:} with 1--10 relevant time-stamped segments.}
}

\endgroup

\section*{Acknowledgements}
We gratefully acknowledge financial support from Mitacs, Ciena, and NSERC of Canada through the Discovery Program. 

\balance
\bibliographystyle{IEEEtran}
\bibliography{ref}

\end{document}